\newcommand{\sla}[1]{#1\!\!\!/}

\documentclass[aps,preprint,showpacs,showkeys]{revtex4}
\usepackage{graphicx}

\begin{document}

\title{Production of Neutral Fermion in Linear Magnetic
Field through Pauli Interaction}

\author{Hyun Kyu \surname{Lee} and Yongsung \surname{Yoon}}

\affiliation{Department of Physics, Hanyang University, Seoul
133-791, Korea}

\begin{abstract}
We calculate the production rate of neutral fermions in linear
magnetic fields through the Pauli interaction.  It is found that
the production rate is exponentially decreasing function with
respect to the inverse of the magnetic field gradient, which shows
the non-perturbative characteristics analogous to the Schwinger
process.  It turns out that the production rate density depends on
both the gradient and the strength of magnetic fields in 3+1
dimension. It is quite different from  the result in 2+1
dimension, where the production rate depends only on the gradient
of the magnetic fields, not on the strength of the magnetic
fields.  It is also found that the  production  of neutral
fermions through the Pauli interaction is a magnetic effect
whereas the production of charged particles through minimal
coupling is an electric effect.
\end{abstract}



\maketitle

\section{INTRODUCTION}

It is well known that the  interaction of  charged spin-1/2
fermions with the electromagnetic field  is described by the
minimal coupling in the form of Dirac equation. Pauli \cite{pauli}
suggested a non-minimal coupling  of spin 1/2-particle with
electromagnetic fields, which can be interpreted as an effective
interaction of fermion to describe the anomalous magnetic moment
of fermions. Pauli interaction is particularly interesting for
describing the interaction of neutral particle with
electromagnetic field provided it has a non-vanishing magnetic
dipole moment \cite{ho}. One of the immediate possibilities
\cite{fujikawa} is the electromagnetic interaction(Pauli
interaction)  of neutrinos, which are recently confirmed to have a
non-zero mass with mixing \cite{valle}. The presence of the
magnetic dipole moment implies that   neutrino can directly couple
to the electromagnetic field, which leads to a variety of new
processes \cite{osc,goyal}.

One of the interesting phenomena with  the strong electromagnetic
field configuration is  the pair creation of particles.   The well
known example is the Schwinger process with the minimal coupling,
in which charged particles are created in pairs \cite{schwinger}
under a strong electric field. However it has been demonstrated
that no particle creation is possible under the pure magnetic
field configuration  even with the spatial inhomogeneity
\cite{dunn}. For a neutral particle with the Pauli interaction,
the inhomogeneity of the magnetic field coupled directly to the
magnetic dipole moment plays an interesting role analogous to the
electric field for a charged particle. The non-zero gradient of
the magnetic field can exert a force on a magnetic dipole moment
such that the neutral fermion can get an energy out of the
magnetic field. Hence it will affects the vacuum structure greatly
for the strong enough magnetic field as for the case of charged
particles in the strong electric field. It is then interesting to
see whether the vacuum production of neutral fermion with a
non-zero magnetic moment in an inhomogeneous magnetic field is
possible on the analogy of the Schwinger process. Interestingly it
has been demonstrated in 2+1 dimension that the magnetic dipole
coupled to the field gradient induces pair creations in a vacuum
\cite{lin}.  In this work, we will present a realistic calculation
in 3+1 dimension for the pair production rate of neutral fermions
with non-vanishing magnetic dipole moment.

\section{Production Rate of Neutral Fermions through Pauli Interaction}

The simplest Lagrangian for the neutral fermion, which couples to
the external electromagnetic field,  was suggested  by Pauli long
time ago \cite{pauli}.  The Dirac equation  with Pauli term is
given by
\begin{equation}
{\cal L} =
\bar{\psi}(\sla{p}+\frac{\mu}{2}\sigma^{\mu\nu}F_{\mu\nu}-m)\psi,\label{pauli}
\end{equation}
where $\sigma^{\mu\nu}=\frac{i}{2}[\gamma^{\mu},\gamma^{\nu}],
~~g_{\mu\nu}=(+,-,-,-)$.  $\mu$ in the Pauli term measures  the
magnitude of the magnetic dipole moment of fermion.  Pauli term
can be considered as an effective interaction which describes
anomalous magnetic moment of fermion or as an effective magnetic
moment induced by the bulk fermions in a theory with large extra
dimensions \cite{ng}. The corresponding Pauli Hamiltonian operator
is
\begin{equation}
H = \vec{\alpha}\cdot
(\vec{p}-i\mu\beta\vec{E})+\beta(m-\mu\vec{\sigma}\cdot\vec{B}),
\label{hamiltonian}
\end{equation}
where $ \sigma^{i}=\frac{1}{2}\epsilon^{ijk}\sigma^{jk}$. If the
magnetic field is stronger than the critical field, $B_c \equiv
m/\mu$, a negative energy state of $m-|\mu B|$ appears. Thus, we
will consider magnetic fields weaker than the critical magnetic
field, $B < B_c $.

In general, the effective potential, $ V_{\rm eff}(A)$,  for a
background electromagnetic
 vector potential, $A_{\mu}$,  can be
obtained by integrating out the fermion:
\begin{eqnarray}
i \int d^{4}x V_{\rm eff}(A[x])
=Tr\ln\{(\sla{p}+\frac{\mu}{2}\sigma^{\mu\nu}F_{\mu\nu}-m)\frac{1}{\sla{p}-m}\},
\end{eqnarray}
where $F_{\mu\nu}=\partial_{\mu}A_{\nu} - \partial_{\nu}A_{\mu}$.
The decay probability of the background magnetic field into the
neutral fermions is related to the imaginary part of the effective
potential $V_{\rm eff}(A)$,
\begin{equation}
P = 1-|e^{i \int d^{4}x  V_{\rm eff}(A[x])}|^{2} = 1-e^{-2Im \int
d^{3}x dt  V_{\rm eff}(A[x])}.
\end{equation}
That is, the twice of the imaginary part of the effective
potential $V_{\rm eff}(A[x])$ is the fermion production rate per
unit volume \cite{field}: $w(x)= 2Im (V_{\rm eff}(A[x]))$.

Using the charge conjugation matrix C:
\begin{equation}
C\gamma_{\mu}C^{-1}=-\gamma^{T}_{\mu}, ~~
C\sigma^{\mu\nu}C^{-1}=-\sigma^{T\mu\nu},
\end{equation}
and  the identity
\begin{equation}
\ln\frac{a}{b}=\int^{\infty}_{0}\frac{ds}{s}(e^{is(b+i\epsilon)}-e^{is(a+i\epsilon)}),
\end{equation}
we can write the effective potential $V_{\rm eff}(A[x])$ as
follows
\begin{equation}
V_{\rm eff}(A[x])=\frac{i}{2} \int^{\infty}_{0}\frac{ds}{s}
e^{-ism^{2}}
tr(<x|e^{is(\sla{p}+\frac{\mu}{2}\sigma^{\mu\nu}F_{\mu\nu})^{2}}|x>-<x|e^{isp^{2}}|x>).
\label{be-w}
\end{equation}

For an inhomogeneous magnetic field, we  consider a static
magnetic field configuration  of $\hat{z}$-direction with a
constant gradient along $\hat{x}$-direction,
$\vec{B}=B(x)\hat{z}$, such that
\begin{equation}
F_{12}=B(x)=B_{0}+B'x=B'\tilde{x},~~ (\tilde{x} = x_{0}+x,
~~x_{0}=\frac{B_{0}}{B'}).
\end{equation}
It is not necessary to consider an infinitely extended
ever-increasing magnetic field to meet the linear magnetic field
configuration. Because the particle production rate density is a
local quantity, it is sufficient to have a uniform gradient
magnetic field in the Compton wavelength scale of the particle.

For this background magnetic field, we get \begin{eqnarray}
(\sla{p}+\frac{\mu}{2}\sigma^{\mu\nu}F_{\mu\nu})^{2}
 =-p_{1}^{2}-p_{2}^{2}+(\mu
B'\tilde{x}+i\tilde{\gamma}^{3}p_{0}+i\tilde{\gamma}^{0}p_{3})^{2}-\mu
B'\gamma^{2}~,
\end{eqnarray}
where $\tilde{\gamma}^{3} \equiv \gamma^{0}\gamma^{1}\gamma^{2}$
and  $~~\tilde{\gamma}^{0} \equiv \gamma^{1}\gamma^{2}\gamma^{3}.$

The second term in  Eq.($\ref{be-w}$) is easily calculated to be
\begin{equation}
v^{(0)} \equiv tr<x|e^{isp^{2}}|x> = -\frac{i}{4\pi^{2}s^{2}}.
\end{equation}
Inserting a complete set of momentum eigenstates, the first term
of Eq.($\ref{be-w}$) can be written by
\begin{eqnarray}
v^{(A)} &\equiv& tr<x|e^{-is\{p_{1}^{2}+p_{2}^{2}-(\mu
B'\tilde{x}+i\tilde{\gamma}^{3}p_{0}+i\tilde{\gamma}^{0}p_{3})^{2}+\mu
B'\gamma^{2}\}}|x> \nonumber \\ ~ &=&
\frac{1}{(2\pi)^{4}}(\frac{\pi}{is})^{\frac{1}{2}} tr\int dp_{1}
dp'_{1} dp_{0} dp_{3}
e^{i(p_{1}-p'_{1})\tilde{x}}<p_{1}|e^{-is(p_{1}^{2}-\mu^{2}B'^{2}x^{2})}|p'_{1}>
\times \nonumber \\ &~& e^{\frac{1}{\mu
B'}(\tilde{\gamma}^{0}p_{3}+\tilde{\gamma}^{3}p_{0})(p'_{1}-p_{1})}
e^{-is\mu B'\gamma^{2}},
\end{eqnarray}
where we have used $[\tilde{\gamma^{3}},\gamma^{2}] = 0 =
[\tilde{\gamma^{0}},\gamma^{2}], ~~[p_{1},\tilde{x}]=-i$ and
\begin{eqnarray}
&p_{1}^{2}+p_{2}^{2}-(\mu
B'\tilde{x}+i\tilde{\gamma}^{3}p_{0}+i\tilde{\gamma}^{0}p_{3})^{2}
= \nonumber \\ &e^{-\frac{1}{\mu
B'}(\tilde{\gamma}^{0}p_{3}+\tilde{\gamma}^{3}p_{0})p_{1}}e^{ix_{0}p_{1}}(p_{1}^{2}+p_{2}^{2}-(\mu
B'x)^{2})e^{\frac{1}{\mu
B'}(\tilde{\gamma}^{0}p_{3}+\tilde{\gamma}^{3}p_{0})p_{1}}e^{-ix_{0}p_{1}}.
\end{eqnarray}

Using the properties of gamma matrices
\begin{equation}
\{\tilde{\gamma^{0}},\tilde{\gamma^{3}}\} =0, ~~~
tr(\tilde{\gamma^{3}}\gamma^{2}) = 0 =
tr(\tilde{\gamma^{0}}\gamma^{2}),
\end{equation}
we get
\begin{equation}
tr e^{\frac{1}{\mu
B'}(\tilde{\gamma}^{0}p_{3}+\tilde{\gamma}^{3}p_{0})(p'_{1}-p_{1})}
e^{-is\mu B'\gamma^{2}} = 4 \cosh(s\mu B')
\cos\{\frac{(p'_{1}-p_{1})}{\mu B'}(p_{0}^{2}-p_{3}^{2})^{1/2}\}.
\end{equation}

Thus, $v^{(A)}$ can be written as follows
\begin{eqnarray}
v^{(A)}&=&\frac{4}{(2\pi)^{4}}(\frac{\pi}{is})^{\frac{1}{2}}
\cosh(s\mu B') \int dp_{1} dp'_{1} dp_{0} dp_{3}
e^{i(p_{1}-p'_{1})\tilde{x}} \times \nonumber \\ &~&
\cos\{\frac{(p_{1}-p'_{1})}{\mu B'}(p_{0}^{2}-p_{3}^{2})^{1/2}\}
<p_{1}|e^{-is(p_{1}^{2}-\mu^{2}B'^{2}x^{2})}|p'_{1}>.  \label{w-A}
\end{eqnarray}

The matrix elements in momentum space in Eq.(\ref{w-A}) correspond
to the matrix elements of the evolution operator for the simple
harmonic oscillator with an imaginary frequency, $ ~\omega=2i\mu
B'$ and $m=\frac{1}{2}$. Then we get
\begin{equation}
<p_{1}|e^{-is(p_{1}^{2}-\mu^{2}B'^{2}x^{2})}|p'_{1}>=\int dx'
dx''<p_{1}|x''>U(x'',s;x',0)<x'|p'_{1}>,
\end{equation}
where
 $U(x,s;x',0)$ is  given by
\begin{equation}
U(x,s;x',0) \equiv <x,s|e^{-is(p_{1}^2+\frac{1}{2}\omega^2 x^2)}
|x',0> =(\frac{\omega}{4\pi i \sin\omega
t})^{1/2}e^{i\omega\frac{(x^{2}+x'^{2})\cos\omega s - 2xx'
}{4\sin\omega s}}. \label{u}
\end{equation}
Performing the $x'$ and $x''$ integration explicitly, we get
\begin{equation}
<p_{1}|e^{-is(p_{1}^{2}-\mu^{2}B'^{2}x^{2})}|p'_{1}>=
(\frac{i\alpha}{\pi})^{\frac{1}{2}} \frac{1}{2\alpha \sin\omega
s}e^{-\frac{i(p_{1}-p'_{1})^{2}}{8\alpha(1-\cos\omega s)}}
e^{\frac{i(p_{1}+p'_{1})^{2}}{8\alpha(1+\cos\omega s)}},
\label{off-diag}
\end{equation}
where $\alpha = -\frac{\omega}{4\sin\omega s}$.

Inserting Eq.($\ref{off-diag}$) into Eq.($\ref{w-A}$), the
integration over $p_{1}$ and $p'_{1}$ gives
\begin{eqnarray}
v^{(A)}&=&
 \frac{2}{(2\pi)^{4}}(\frac{\pi}{is})^{\frac{1}{2}}\coth(s\mu
B')
{\cal K},
\end{eqnarray}
where, defining $\gamma, a$ and $p_{-}$ as follow
\begin{equation}
 \gamma \equiv \frac{\coth(s\mu B')}{4\mu B'}, ~~~
 a \equiv \frac{1}{\mu
B'}(p_{0}^{2}-p_{3}^{2})^{\frac{1}{2}}, ~~~p_ = p_{1}-p'_{1},
\end{equation}
${\cal K}$ is given by
\begin{eqnarray}
{\cal K} &\equiv& \frac{1}{2}
\int^{\infty}_{-\infty}dp_{-}dp_{0}dp_{3}
e^{i\tilde{x}p_{-}}(e^{iap_{-}}+e^{-iap_{-}})e^{-i\gamma
p_{-}^{2}}, \nonumber
\\&=&\frac{1}{2}(\frac{\pi}{i\gamma})^{1/2}\int
dp_{0}dp_{3}(e^{i\frac{(a+\tilde{x})^{2}}{4\gamma}}+e^{i\frac{(a-\tilde{x})^{2}}{4\gamma}})
\nonumber \\
&=& 4\pi(\mu
B')^{2}(\frac{\pi\gamma}{i})^{\frac{1}{2}} - 2(\mu B)^{2}i \pi
(\frac{\pi}{i\gamma})^{\frac{1}{2}}\int^{1}_{0}d\xi(1-\xi)e^{i\frac{\tilde{x}^{2}}{4\gamma}
\xi^{2}}.
\end{eqnarray}

Thus, $v^{(A)}$ is reduced to the following
\begin{equation}
v^{(A)}=-\frac{1}{4\pi^{2}}[\frac{i}{s^{2}}\{(s\mu B')\coth(s\mu
B') \}^{\frac{3}{2}} +2(\mu B)^{2}\frac{1}{s}\{(s\mu B')\coth(s\mu
B')\}^{\frac{1}{2}}\int^{1}_{0}d\xi
(1-\xi)e^{i\frac{\tilde{x}^{2}}{4\gamma} \xi^{2}}]
\end{equation}
and finally we get  the effective potential $V_{\rm eff}$ as
follows
\begin{equation}
V_{\rm eff}=\frac{i}{2}
\int^{\infty}_{0}\frac{ds}{s}e^{-ism^{2}}(v^{(A)}-v^{(0)}).
\label{Veff}
\end{equation}

The effective potential for the uniform field configuration can be
obtained by putting  $\mu B' =0$ to get
\begin{equation}
V_{\rm eff}=-\frac{(\mu B)^{2}}{4\pi^{2}}
\int^{\infty}_{0}\frac{ds}{s^{2}}\{i\int^{1}_{0}d\xi(1-\xi)e^{i(\mu
B)^{2} \xi^{2}s}-\frac{i}{2}+\frac{(\mu B)^{2}s}{12} \}
e^{-im^{2}s}. \label{VC}
\end{equation}
The divergent contributions at $s=0$ are removed by adding local
counter terms of $(\mu B)^{2}$, and $(\mu B)^{4}$. This implies
the renormalization of the magnetic moment $\mu$ to the measured
value and the coupling of $(\mu B)^{4}$ to zero presumably. The
effective potential, Eq.($\ref{VC}$), for uniform magnetic fields
is found to be real, which  implies  a stable magnetic background
of $w=0$. For magnetic fields weaker than the critical field
$B_c=m/\mu$, using a contour integration in the fourth quadrant,
the integration can be done along the negative imaginary axis
giving the finite real effective action as
\begin{equation} V_{\rm eff}=\frac{(\mu B)^{2}}{4\pi^{2}}
\int^{\infty}_{0}\frac{ds}{s^{2}}\{\frac{1}{2}+\frac{(\mu
B)^{2}s}{12}-\int^{1}_{0}d\xi(1-\xi)e^{(\mu B)^{2} \xi^{2}s} \}
e^{-m^{2}s}. \label{VCR}
\end{equation}
The leading radiative correction term for a weak $B$ field is
\begin{equation}
\delta V_{\rm eff}=\frac{(\mu B)^{6}}{240\pi^{2}m^{2}}.
\end{equation}

For an inhomogeneous field configuration, $\mu B' \neq 0$, the
effective potential is given by
\begin{eqnarray}
V_{\rm eff} &=& -\frac{(\mu B)^{2}}{4\pi^{2}}
\int^{\infty}_{0}\frac{ds}{s^{2}}\{ i\sqrt{\mu B's\coth(\mu B's)}
\int^{1}_{0} d\xi(1-\xi) e^{i\frac{(\mu B)^{2}}{\mu B'
}\xi^{2}\tanh(\mu B's)}-\frac{i}{2}+\frac{(\mu B)^{2}s}{12}
\}e^{-im^{2}s} \nonumber
\\ &~&
+\frac{1}{8\pi^{2}}\int^{\infty}_{0} \frac{ds}{s^{3}} \{ (\mu
B's\coth (\mu B's))^{3/2}-1-\frac{(\mu B's)^{2}}{2} \}
e^{-im^{2}s}, \label{VG}
\end{eqnarray}
where an additional divergent contribution at $s=0$ is removed by
adding a local counter term of $(\mu B')^{2}$ in the second term.

The leading radiative correction terms of the effective potential
for a small gradient weak $B$ field are calculated as given by
\begin{equation}
\delta V_{\rm eff}=\frac{(\mu B)^{6}}{240\pi^{2}m^{2}} +\frac{(\mu
B)^{4}(\mu B')^{2}}{288\pi^{2}m^{4}} -\frac{(\mu B)^{2}(\mu
B')^{2}}{48\pi^{2}m^{2}} -\frac{(\mu B')^{4}}{960\pi^{2}m^{4}}.
\end{equation}

It is found that the effective potential, Eq.($\ref{VG}$), has a
non-vanishing imaginary part, which implies that the background of
inhomogeneous magnetic field configuration is unstable against the
creation of neutral fermions with Pauli interaction.  From the
imaginary part of the effective potential Eq.($\ref{VG}$), we
obtain the production rate density, $w(x)=2Im(V_{\rm eff}(A[x]))$.

Introducing dimensionless parameters defined as  $v = s\mu B',~~
\lambda = \frac{m^{2}}{|\mu B'|},~~ \kappa = \frac{m^{2}}{(\mu
B)^{2}}$, the production rate density $w(x)$ in the unit of the
fermion mass is finally given by
\begin{eqnarray}
w(x) =
&-&\frac{2m^{4}}{4\pi^{2}\lambda\kappa}\int^{\infty}_{0}\frac{dv}{v^{2}}
\{ \sqrt{v\coth v} F(\frac{\lambda}{\kappa}\tanh v,\lambda v)
-\frac{1}{2}\cos\lambda v -\frac{\lambda v}{12\kappa}\sin\lambda v
\} \nonumber
\\
&-&\frac{m^{4}}{4\pi^{2}\lambda^{2}}\int^{\infty}_{0}\frac{dv}{v^{3}}\{(v
\coth v)^{3/2}-1-\frac{v^{2}}{2} \}\sin \lambda v, \label{w}
\end{eqnarray}
where
\begin{eqnarray}
F(a,b) &\equiv& \int^{1}_{0}d\xi (1-\xi) \cos(a\xi^{2}-b)
\nonumber
\\ &=& -\frac{1}{2a}\{ \sin (a - b) + \sin (b) \} \nonumber \\
&~& + \sqrt{\frac{\pi }{2a}}\ \{ \cos (b)\,{\rm
FresnelC}(\sqrt{\frac{2a}{\pi }}) +
        \sin (b)\,{\rm FresnelS}(\sqrt{\frac{2a}{\pi }}) \}.
\end{eqnarray}
Since the scale of inhomogeneity less than Compton wavelength of
the fermion is irrelevant to the particle production through this
process, we take in this work  the spatial gradient of the
magnetic field $|B'|$ to be smaller than the ratio of field
strength, $|B|$, to the Compton wavelength $\frac{1}{m}$, that is,
$\lambda > \sqrt{\kappa}$.

The integration Eq.($\ref{w}$) is a finite integration, but has
singularities along the imaginary $v$ axis similarly to the
Schwinger's result, where the residue calculation gives the
analytic WKB type expression. However, the integration
Eq.($\ref{w}$) has essential sigularities along the imaginary
axis, so that it seems not possible to get a usual analytic WKB
type expression using a contour integration. Therefore we use
numerical integrations to investigate the properties of the
production rate density given by Eq.($\ref{w}$).  For the
numerical calculation, we consider the case of  $\kappa \geq 1$
and $\lambda > 2$ as an example. Numerical integrations of the
production rate density show that the second term of
Eq.($\ref{w}$) is negligible compared to the first term for
$\lambda > 2$.   The production rate shows exponential monotonic
decrease for $\lambda > 2\sqrt{\kappa}$ and $\kappa \geq 1$.   The
production rate, $w(x)$, is calculated for $\kappa=1.0$ and $2.0$
as a function of  $\lambda$. The results in the unit of
$\frac{m^{4}}{4\pi^{2}}$ are  as shown in [Fig.1-2].  The results
of numerical calculations are represented by dots in the figures.
The numerical integrations of Eq.($\ref{w}$) suffer from large
oscillatory fluctuation, which is an unavoidable feature due to
the violent oscillations in the integrand as discussed in
\cite{grifols}. To get an analytic expression, we obtain the best
fit of the numerical results to the curves in the form of
$\frac{a_{\kappa}}{\lambda}e^{-b_{\kappa}\lambda}$ for [Fig.1-2].
We can observe that the particle creation rate is an exponentially
decreasing function with respect to the inverse of the field
gradient,
\begin{eqnarray}
w \sim
 e^{- {\rm constant} \times m^2/|\mu B'|}, \label{wexp}
\end{eqnarray}
which shows the  characteristics of the non-perturbative process.
This can be understood as a quantum tunnelling  through a
potential barrier of height $\sim 2m$ of a particle exposed to an
potential energy $ \sim \mu |B'| x $ due to the inhomogeneous
magnetic field coupled to the magnetic dipole moment  through
Pauli interaction. It is similar to the Schwinger process  of
electron-positron pair creation in the  strong electric field,
where the creation rate is decreasing exponentially \cite{field},
$w \sim e^{- {\rm constant} \times m^2/|e E|}$.

One can see  that the production rate is suppressed very rapidly
when the field strength becomes weaker than $\sim m/\mu$ as well
as the inhomogeneity scale is bigger than Compton wave length
scale. For the inhomogeneity in the Compton wave length scale, the
rate per unit time per unit volume is typically of order  $
\frac{m^4}{4\pi^2} $ for the critical field strength.

However for the realistic estimation of the production rate, more
precise information on the mass and magnetic moment of a particle
and the strength of the magnetic field  as well as the scale of
the spatial inhomogeneity of the field  in consideration are
needed for the observational possibility. As a possible
environment, let us consider the pair creation of neutrino with  a
non-zero magnetic dipole moment in the vicinity of   the very
strongly magnetized compact objects with $B = 10^{15}$G as a
typical strength \cite{magnetar,inside}.  Taking the possible
magnetic moment to be as large as the experimental upper bound
$\mu_{\nu}=10^{-11}\mu_B$ and the mass of the neutrino to be
$m_{\nu} \sim 10^{-2}$ eV constrained by the solar neutrino
observations, the critical field strength is estimated to be  $B_c
\sim 10^{17}$ G.   One can see that the condition for $\kappa \geq
1$ or $B < B_c$ assumed in this work is satisfied. Since the the
scale of the inhomogeneity is  naturally about the size of the
compact object, $R \sim 10^4$ m, which   is much larger than the
Compton wave length $\sim 10^{-4}$m, the production rate is
expected to be substantially suppressed from the typical rate such
that  it may not provide sufficiently high luminosity for the
neutrino detectors.

\section{Discussion}

We have examined  the vacuum production of neutral fermions in
inhomogeneous magnetic fields through Pauli interaction.  The
fermions, which are coupled to the background electromagnetic
field through Pauli interaction,  are integrated out and  there
appears an imaginary part in the effective action. It turns out
that the production rate density depends on both the gradient and
the strength of magnetic fields in 3+1 dimension, which is quite
different from the result in 2+1 dimension \cite{lin}, where the
production rate depends only on the gradient of the magnetic
fields, not on the strength of the magnetic fields. The difference
can be attributed to the different nature  of spinors in 3+1 and
2+1 dimensions.

The vacuum production of fermion with the Pauli interaction is
found to be a magnetic effect.  Explicit calculations with a
linear electric field configuration of $\hat{x}-$direction with a
constant gradient along $\hat{x}-$direction such that
$E_{x}=E_{0}+E'x$ shows  that the effective potential has no
imaginary part when the singularities are regularized properly. It
can be also shown by substituting $B \rightarrow iE$ and $B'
\rightarrow iE'$ in the effective potential for a pure magnetic
field Eq.($\ref{Veff}$) with the $s$ integration along the
imaginary axis. Therefore one can see that the pair creation
through Pauli interaction is a purely magnetic effect. It is an
interesting result when compared to the pair creation of charged
particles through the minimal coupling, which is known to be an
electric effect \cite{field}.

Although the production rate density in this work has been derived
for $\mu B'= {\rm constant}$, it can be applicable to various
types of magnetic fields provided that the magnetic field is
linear in the scale of  Compton wavelength  of the particle
considered because the particle production rate density is a local
quantity. It may be therefore  applicable to a spatially slowly
varying $\mu B'(x)$ as a good approximation if the gradient
variation is very small in the Compton wavelength scale.  For the
realistic calculation of the production rate, more precise
information on the mass and magnetic moment of  a particle and the
strength of the magnetic field  as well as the scale of the
spatial inhomogeneity of the field in consideration are needed for
the observational possibility.

This work was supported  by Korea Research Foundation
Grant(KRF-2004-041-C00085).


\begin{figure}[htp]
\begin{center}
\includegraphics[height=2.5in,width=2.5in]{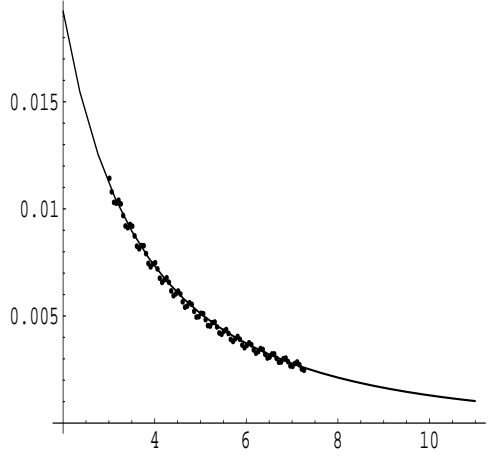}
\caption{ $\kappa=1.0$ with varying $\lambda$: $a_{1.0}=0.050,~
b_{1.0}=0.136$~}
\end{center}
\end{figure}

\begin{figure}[htp]
\begin{center}
\includegraphics[height=2.5in,width=2.5in]{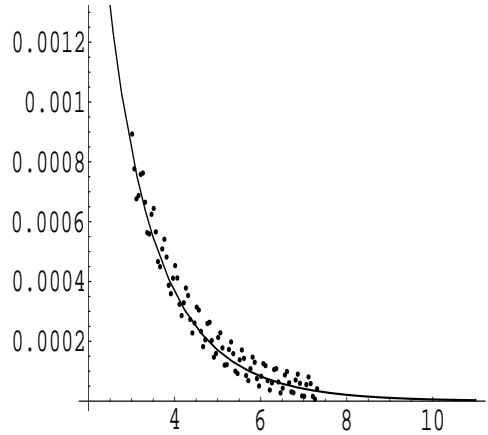}
\caption{ $\kappa=2.0$ with varying $\lambda$: $a_{2.0}=0.013,~
b_{2.0}=0.775~$}
\end{center}
\end{figure}


\begin{references}

\bibitem{pauli} W. Pauli, Rev. Mod. Phys. 13, 203(1941)

\bibitem{ho} See, for example, C-L. Ho and P. Roy,  Annals. Phys. 312,
161(2004); Q-q Lin, Phys. Rev. A61, 022101(2000)

\bibitem{valle} For a recent review see, {\it e.g.},  J. F. Valle,
hep-ph/0508067(2005)


\bibitem{fujikawa} K. Fuzikawa and R. Shrock, Phys. Rev. Lett.
45, 963(1980)

\bibitem{osc} E. K. Akhmedov, Phys. Lett. B213, 64(1988);C.-S. Lim
and J. Marciano, Phys. Rev. D37, 1368(1988) ;A.B. Balantekin and
C. Volpe, Phys. Rev. D72, 033008(2005)

\bibitem{goyal} A. Goyal, Phys. Rev. D64, 013005(2001)

\bibitem{schwinger} J. Schwinger, Phys. Rev. 82, 664(1951)

\bibitem{dunn} G. Dunne,  Phys.Lett. B419,  322(1998); Phys. Rev.
D60, 065002(1999)

\bibitem{lin} Q-G Lin, J. Phys. G. 25, 1793(1999)


\bibitem{ng} G.C. McLaughlin and J.N. Ng, Phys. Lett. B470, 157(1999)

\bibitem{field} C. Itzykson and J.B. Zuber, {\it Quantum Field
Theory}(McGraw-Hill, New York, 1980)

\bibitem{grifols} J.A. Grifols and E. Mass$\acute{o}$, Phys. Rev. D65,
055004 (2002)

\bibitem{magnetar}  A. K. Harding, astro-ph/0510134(2005)

\bibitem{inside} R.C. Duncan and C. Thompson, ApJ 392, 9(1992)


\end{references}
\end{document}